\documentclass[a4paper]{jpconf}
\usepackage{graphicx}
\usepackage{amssymb}
\begin{document}
\title{Towards the (Mexican) discovery of second class currents at Belle-II}

\author{Pablo Roig}

\address{Departamento de F\'isica, Centro de Investigaci\'on y de Estudios Avanzados del Instituto Polit\'ecnico Nacional, Apartado Postal 14-740, 07000 M\'exico D.F., M\'exico}

\ead{proig@fis.cinvestav.mx}

\begin{abstract}
Within the SM, the yet unmeasured $\tau^-\to\pi^-\eta^{(\prime)}\nu_\tau$ decays are predicted as a suppressed, isospin-violating effect with branching ratios $\lesssim\mathcal{O}(10^{-5})$. 
However, they can also proceed through other mechanisms (such as charged Higgs exchange) at comparable rates. This has motivated several studies of the corresponding QCD predictions 
for these second class current processes. In this contribution we discuss the prospects for the discovery of these decays at Belle-II emphasizing the Mexican involvement in this project. 
Our branching ratio prediction $\sim1.7\cdot10^{-5}$ (decay channel with an $\eta$ meson) is well within the reach of Belle-II. The branching fraction for the decay channel with an 
$\eta^\prime$ meson is expected to be between one and two orders of magnitude more suppressed.
\end{abstract}

\section{Introduction}
Non-strange weak left-handed hadron currents \footnote{We note that this classification is only determined by the quantum numbers of the final-state hadrons, being therefore independent of 
possible (different) intermediate resonance exchange mechanisms.} are split according to their spin ($J$), parity ($P$) and $G$-parity ($G$) \cite{Weinberg:1958ut} as:\\
i) first class currents, with $J^{PG}=0^{++},0^{--},1^{+-},1^{-+}$;\\
ii) second class currents (SCC), which have opposite $G$-parity to those considered above.\\
Within the SM, $\tau^-\to\pi^-\eta^{(\prime)}\nu_\tau$ decays can only arise through SCC because the $S(P)$-waves have $J^P=0^+(1^-)$ but the $\pi\eta^{(\prime)}$ system has $G=-$. However, they are 
\textit{induced} by a small isospin violating parameter $\varepsilon_{\pi\eta^{(\prime)}}\sim\frac{m_u-m_d}{m_s}\sim10^{-2}$ at the amplitude level, which is fixed by chiral symmetry alone 
for the $SU(3)$ flavor octet $\eta_8$ state \cite{Gasser:1984gg}. 
This isospin (equivalently $G$-parity) breaking can be understood as the allowed production of the $\pi^-\pi^0$ system \footnote{In fact it is the dominant decay mode of the $\tau$ lepton with a 
branching fraction of $\sim25\%$.}  with a subsequent $\pi^0-\eta^{(\prime)}$ mixing regulated by $\varepsilon_{\pi\eta^{(\prime)}}$. 
On the contrary, new physics contributions can yield enhanced rates if \textit{genuine} SCC do exist.\\
Shortly after the $\tau$ lepton discovery, it was pointed out \cite{Leroy:1977pq} that the search for $\tau^-\to\pi^-\eta\nu_\tau$ decays (a prominent scalar contribution from 
$a_0(980)$ meson intermediate exchange was already noticed) could lead to the discovery of SCC.\\
Interest in SCC was renewed ten years later, when the HRS Coll. \cite{Derrick:1987sp} announced $BR(\tau^-\to\pi^-\eta\nu_\tau)\sim\mathcal{O}(5\%)$, against the expected suppression given 
by the overall $\varepsilon_{\pi\eta}^2$ factor, which would yield a branching fraction at the level of $10^{-4}$, at most. The situation settled quickly, with the SM prediction of $\tau$ 
decays to final states including $\eta$ mesons \cite{Pich:1987qq} evidencing some issue in the HRS data, in agreement with other analyses of the $\tau^-\to\pi^-\eta\nu_\tau$ decays 
\cite{Berger:1987sj, Bramon:1987zb, Braaten:1989zn}, including Mexican contributions already \cite{DiazCruz:1991se}~\footnote{Incidentally, one of the authors of Ref.~\cite{Leroy:1977pq} (J.~P.) was the 
Ph. D. thesis advisor of one of the authors of Ref.~\cite{DiazCruz:1991se} (G.~L.~C.). The other author of the last reference (J.~L.~D.~C.) was a postdoc with one of the authors of 
Ref.~\cite{Bramon:1987zb} (A.~B.). These precedents probably gave rise to the early Mexican interest in SCC.}. Branching ratios in the range 
$\left[1.2,1.8\right]\cdot10^{-5}$ were predicted by that time \cite{Bramon:1987zb, Tisserant:1982fc, Neufeld:1994eg} for this decay mode.\\
Current bounds on these decays were set by the BaBar Coll.: $BR(\tau^-\to\pi^-\eta\nu_\tau)<9.9\cdot 10^{-5}$ \cite{delAmoSanchez:2010pc} and $BR(\tau^-\to\pi^-\eta^\prime\nu_\tau)<4.0\cdot 
10^{-6}$ \cite{Lees:2012ks}, at $95$ and $90\%$ C.L., respectively. Belle-II is expected to accumulate up to two orders of magnitude more data than BaBar and Belle \cite{Abe:2010gxa}, 
which should make possible the discovery of SCC. Elucidating whether these are of the expected \textit{induced} type within the SM, or of \textit{genuine} kind (new physics contributions) 
motivates reconsidering the QCD prediction for these decays.\\
In section \ref{FFs} we introduce the two form factors (scalar and vector) that encode the underlying dynamics of the considered SCC. While we find that the vector one can be extracted from data in a 
model-independent way, this is not possible for the scalar one, for which a fully analytic and unitary treatment is needed, as we discuss. Based on this hadronic input, we give our predictions in section 
\ref{Pred} and state our conclusions and give an outlook in section \ref{Outlook}.\\

\section{Vector and scalar form factors}\label{FFs}

\begin{figure}[h!]
\begin{center}
\includegraphics[scale=0.75]{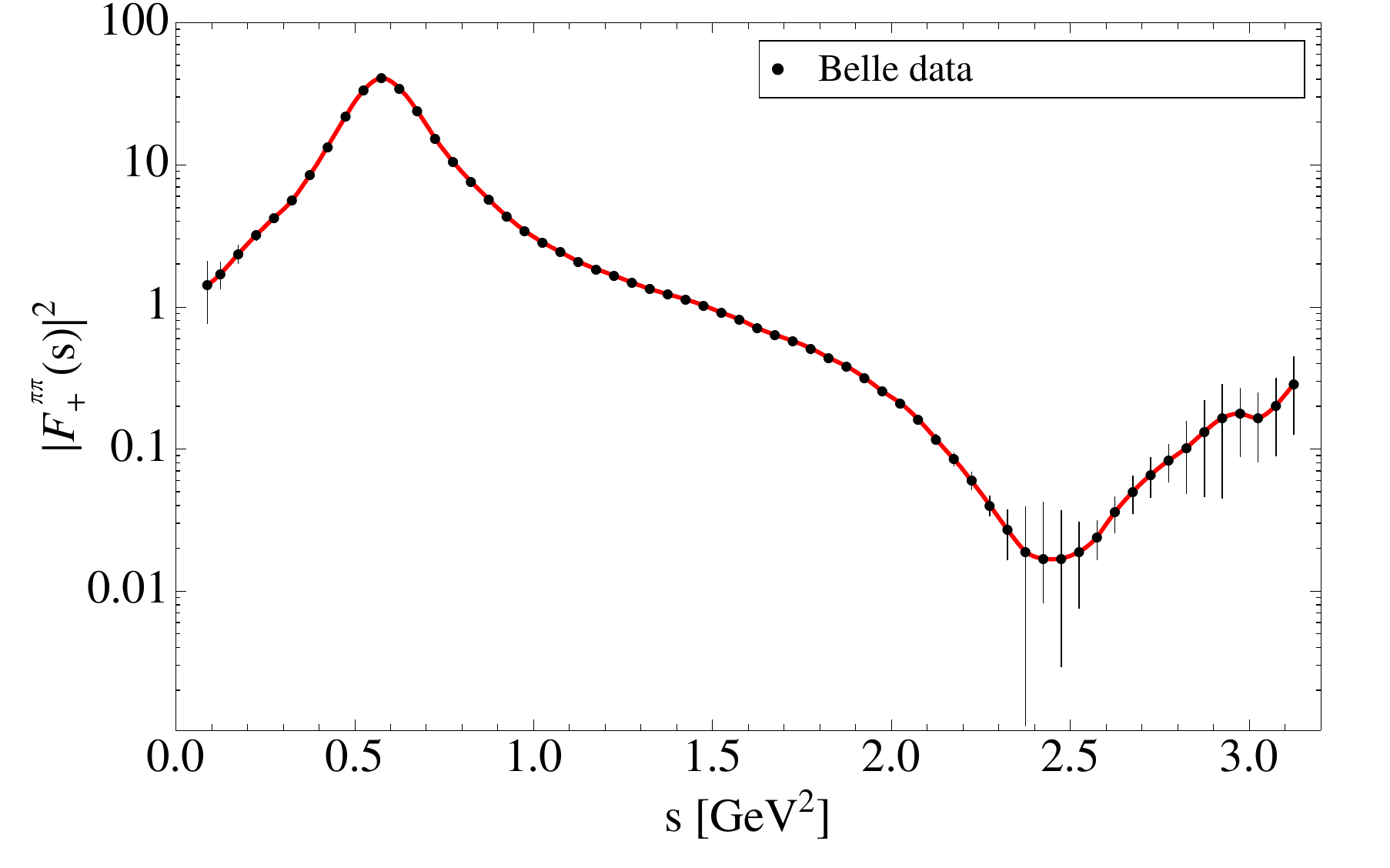}
\caption{\label{VFFpipi} Vector form factor of the $\pi^{-}\pi^{0}$ as obtained in Ref.~\cite{Fujikawa:2008ma} (black circles). The red solid curve corresponds to our data interpolation used in this work.}
\end{center}
\end{figure}

From one point of view, semileptonic $\tau$ decays allow for a clean study of the hadronization of QCD currents in the GeV region \cite{Pich:2013lsa}, since the electroweak half of the 
process is well known theoretically. On the other hand, this hadronic input limits the accuracy of the SM predictions on exclusive meson $\tau$ decays, like those considered here. The 
hadronic matrix element depends on two form factors, $F_{0}^{\pi^-\eta^{(\prime)}}(s)$ $\left(F_{+}^{\pi^-\eta^{(\prime)}}(s)\right)$, carrying $J^P=0^+(1^-)$, respectively:
\begin{equation}
\langle \pi^{-}\eta^{(\prime)}|\bar{d}\gamma^{\mu}u|0\rangle=\left[(p_{\eta^{(\prime)}}-p_{\pi})^{\mu}+\frac{\Delta_{\pi^{-}\eta^{(\prime)}}}{s}q^{\mu}\right]c^{V}_{\pi\eta^{(\prime)}}F_{+}^{\pi^-\eta^{(\prime)}}(s)+\frac{\Delta^{QCD}_{K^{0}K^{+}}}{s}q^{\mu}c^{S}_{\pi^{-}\eta^{(\prime)}}F_{0}^{\pi^{-}\eta^{(\prime)}}(s)\,,
\label{vectorcurrent}
\end{equation}
where \cite{Escribano:2016ntp} $c^{V}_{\pi^{-}\eta^{(\prime)}}=\sqrt{2}$, $s=q^{2}=(p_{\eta^{(\prime)}}+p_{\pi^{-}})^{2}$, $c_{\pi^{-}\eta}^{S}=\sqrt{\frac{2}{3}}$, 
$c_{\pi^{-}\eta^{\prime}}^{S}=\frac{2}{\sqrt{3}}$, $\Delta_{PQ}=m_{P}^{2}-m_{Q}^{2}$ and the superscript $QCD$ indicates that the $K^0K^+$ electromagnetic mass difference does not 
contribute to eq.~(\ref{vectorcurrent}). It is also worth noting that the finiteness of the matrix element (\ref{vectorcurrent}) relates the two form factors at the origin:
\begin{equation}
 F_+^{\pi^-\eta^{(\prime)}}(0)=-\frac{c^S_{\pi^-\eta^{(\prime)}}}{c^V_{\pi^-\eta^{(\prime)}}}\frac{\Delta^{QCD}_{K^0K^+}}{\Delta_{\pi^-\eta^{(\prime)}}}F_0^{\pi^-\eta^{(\prime)}}(0)\,.
\label{FF0} 
\end{equation}
Then, the overall suppression of these decays is given by $F_{+}^{\pi^{-}\eta^{(\prime)}}(0)\sim\mathcal{O}(m_{d}-m_{u})$ \footnote{In fact $F_{+}^{\pi^{-}\eta}(0)\sim\mathcal{O}(\varepsilon_{\pi\eta})$, 
while an accidental cancellation makes $F_{+}^{\pi^{-}\eta^\prime}(0)\sim\mathcal{O}(\varepsilon_{\pi\eta}^2)$, see eq.~(\ref{VFF0}).}. In order to display this neatly, the reduced form 
factors 
\begin{equation}
\widetilde{F}_{+,0}^{\pi^-\eta^{(\prime)}}(s)=\frac{F_{+,0}^{\pi^-\eta^{(\prime)}}(s)}{F_{+,0}^{\pi^-\eta^{(\prime)}}(0)}
\end{equation}
will be used. In fact, as first put forward in Ref.~\cite{Escribano:2016ntp}, the relation
\begin{equation}
\widetilde{F}_{+}^{\pi^{-}\eta}(s)=\widetilde{F}_{+}^{\pi^{-}\eta^{\prime}}(s)=\widetilde{F}_{+}^{\pi^{-}\pi^{0}}(s)\,
\label{VFFnormalized}
\end{equation}
allows a data-driven extraction of $\widetilde{F}_{+}^{\pi^-\eta^{(\prime)}}(s)$ using the charged pion vector form factor, which has been measured precisely by the Belle Coll. 
\cite{Fujikawa:2008ma}, as shown in figure \ref{VFFpipi}.
The values of the form factors at the origin are fixed in terms of the mixing parameters of the $\pi^0-\eta-\eta'$ system ($\varepsilon_{\pi\eta}$, $\varepsilon_{\pi\eta^\prime}$ and 
$\theta_{\eta\eta'}$). At next-to-leading order in Chiral Perturbation Theory \cite{ChPT} in the large-$N_C$ limit \cite{largeNC, largeNCChPT}, this reads \cite{Escribano:2016ntp}
\begin{equation}
F_{+}^{\pi^{-}\eta}(0)=\varepsilon_{\pi\eta}\,,\quad F_{+}^{\pi^{-}\eta^{\prime}}(0)=\varepsilon_{\pi\eta^{\prime}}\,,
\label{VFF0}
\end{equation}
and $\theta_{\eta\eta'}=-(13.3\pm0.5)^{\circ}$~\cite{Ambrosino:2006gk}, $\varepsilon_{\pi\eta}=(9.8\pm0.3)\cdot10^{-3}$, $\varepsilon_{\pi\eta^{\prime}}=(2.5\pm1.5)\cdot10^{-4}$~\cite{Kroll:2005sd}.\\

According to the previous discussion, the vector form factor contribution to $\tau^-\to\pi^-\eta^{(\prime)}\nu_\tau$ decays is fixed with only a small theory error coming basically from 
the uncertainty on $\varepsilon_{\pi\eta^{(\prime)}}$. Unfortunately, the situation is opposite for the scalar form factor contribution. In this case there is no model-independent method for 
extracting $\widetilde{F}_{0}^{\pi^-\eta^{(\prime)}}(s)$. In Ref.\cite{Escribano:2016ntp} several approaches have been considered to obtain it. These start from the most basic Breit-Wigner-like 
construction (which violates analyticity maximally) and goes on considering different approximations with increasing fulfillment of unitarity and analyticity constraints (see also 
\cite{Jamin:2001zq, Escribano:2013bca, Escribano:2014joa} for a thorough discussion of these issues in the case of strangeness-changing scalar form factors). Here we will only discuss 
the most satisfactory approach, which considers the three-coupled channel problem ($\pi^-\eta$, $K^-K^0$, $\pi^-\eta^\prime$) in a dispersive treatment (thus fulfilling analyticity by 
construction) which unitarizes the relevant meson-meson scattering amplitudes obtained within the framework of $U(3)$ Chiral Perturbation Theory including resonances \cite{Guo:2011pa}.

In the elastic region, a two-meson form factor is related to the corresponding meson-meson scattering amplitude via the optical theorem. For our particular case, this reads
\begin{equation}
{\rm{Im}}F_{0}^{\pi^{-}\eta^{(\prime)}}(s)=\sigma_{\pi^{-}\eta^{(\prime)}}(s)F_{0}^{\pi^{-}\eta^{(\prime)}}(s)t_{1,0}^{\pi^{-}\eta^{(\prime)}*}(s)\,,
\label{unitarityrelation}
\end{equation}
where $t_{1,0}^{\pi^{-}\eta^{(\prime)}}(s)$ is the unitarized elastic $\pi^{-}\eta^{(\prime)}$ partial wave of the scattering amplitude of $I=1$ and $J=0$ and 
$\sigma_{\pi^{-}\eta^{(\prime)}}(s)$ is the two-particle phase space function. In Ref.~\cite{Guo:2011pa} the relevant meson-meson scattering amplitudes have been properly unitarized 
through the $N/D$ method \cite{Oller:1998zr,Oller:2000ma}, allowing for the simplified perturbative solution
\begin{equation}
t_{1,0}^{\pi^{-}\eta^{(\prime)}}(s)=\frac{\sigma_{\pi^{-}\eta^{(\prime)}}(s)N^{\pi^{-}\eta^{(\prime)}}_{1,0}(s)}{\left(1+g^{\pi^{-}\eta^{(\prime)}}(s)N^{\pi^{-}\eta^{(\prime)}}_{1,0}(s)\right)}\,,
\label{NDmethod}
\end{equation}
where $g^{\pi^{-}\eta^{(\prime)}}(s)$ are the $\pi^{-}\eta^{(\prime)}$ one-loop scalar functions defined in Eq.~(33) of Ref.~\cite{Guo:2011pa} and $N^{\pi^{-}\eta^{(\prime)}}_{1,0}(s)$ 
contain the expressions of the partial wave amplitudes up to next-to-next-to-leading order in the chiral counting in the large-$N_C$ limit. The central values of the low-energy constants and resonance 
parameters are obtained from Ref.~\cite{Guo:2012yt}. However, our ignorance on the correlations between the errors of the different parameters forbids us to give precise estimates of our uncertainties 
in section \ref{Pred}.\\
In the elastic region the form factor can, of course, be built using the Omn\`es dispersive representation \cite{Omnes:1958hv}. However, from the computational point of view, it is more 
convenient to construct it employing the closed-form solution \cite{Jamin:2001zq, Babelon:1976kv}
\begin{equation}
\widetilde{F}_{0}^{\pi^{-}\eta^{(\prime)}}(s)=\prod_{i=1}\frac{1}{\left(s-s_{p}^{i}\right)\left(1+g^{\pi^{-}\eta^{(\prime)}}(s)N^{\pi^{-}\eta^{(\prime)}}_{1,0}(s)\right)},
\label{elasticclosed}
\end{equation}
where the pre-factors $(s-s_{p}^i)^{-1}$ cancel possible poles of eq.~(\ref{elasticclosed}), which have to be removed in $\widetilde{F}_{0}^{\pi^{-}\eta^{(\prime)}}(s)$. In our 
specific case, the pole resides at $s_{p}=1.9516$ GeV$^2$ corresponding to the bare (squared) mass of the scalar octet \cite{Guo:2012yt}.

A closed-form solution becomes even more convenient in the inelastic region, where unitarity demands a coupled-channel treatment of the problem. In Ref. \cite{Escribano:2016ntp} the two coupled-channel 
$\pi\eta-\pi\eta^\prime$ problem is considered as a starting point. However we find that the $K^-K^0$ contribution needs to be accounted for in order to obtain accurate $\widetilde{F}_{0}^{\pi^{-}\eta^{(\prime)}}(s)$ 
form factors~\footnote{On the contrary, the $\pi^-\pi^0$ contribution is negligible in both cases. This happens because resonances cannot contribute in this channel at first order in isospin breaking.}. In 
this case,
\begin{eqnarray}
g_{1,0}(s)=\left(
\begin{array}{ccc}
             g_{\pi^{-}\eta}(s) & 0&0 \\
             0&g_{K^-K^0}(s)&0\\
             0&0 & g_{\pi^{-}\eta^{'}(s)}
\end{array}\right)
\end{eqnarray}
and
\begin{eqnarray}
N_{1,0}(s)=\left(
\begin{array}{ccc}
             N_{\pi^{-}\eta\to\pi^{-}\eta}(s) & N_{\pi^{-}\eta\to K^{-}K^{0}}(s) & N_{\pi^{-}\eta\to\pi^{-}\eta^{\prime}}(s)\\
             N_{K^{-}K^{0}\to\pi^{-}\eta}(s) & N_{K^{-}K^{0}\to K^{-}K^{0}}(s) & N_{K^{-}K^{0}\to\pi^{-}\eta^{\prime}}(s) \\
             N_{\pi^{-}\eta^{\prime}\to\pi^{-}\eta}(s) & N_{\pi^{-}\eta^{\prime}\to K^{-}K^{0}}(s) & N_{\pi^{-}\eta^{\prime}\to\pi^{-}\eta^{\prime}}(s) 
\end{array}\right)
\end{eqnarray}
encode the corresponding scalar loop functions and partial-wave amplitudes, which enter the master formula for the determination of the participant scalar form factors
\begin{eqnarray}\left(
\begin{array}{c}
	F_{0}^{\pi^{-}\eta}(s)\\
	F_{0}^{K^{-}K^0}(s)\\
	F_{0}^{\pi^{-}\eta^{\prime}}(s)
\end{array}\right)=\frac{1}{Det[D(s)]}
D^{-1}(s)
\left(
\begin{array}{c}
	F_{0}^{\pi^{-}\eta}(0)\\
	F_{0}^{K^{-}K^0}(0)\\
	F_{0}^{\pi^{-}\eta^{\prime}}(0)
\end{array}\,\right),
\label{3ccmatrix}
\end{eqnarray}
where
\begin{equation}
 D(s)\,=\,\left(
\begin{array}{ccc}
 1+g_{\pi^-\eta}(s)T_{\pi^-\eta\to\pi^-\eta}(s) & g_{\pi^-\eta}(s)T_{\pi^-\eta\to K^-K^0}(s) & g_{\pi^-\eta}(s)T_{\pi^-\eta\to\pi^-\eta^\prime}(s)\\
 g_{K^-K^0}(s)T_{K^-K^0\to\pi^-\eta}(s) & 1+g_{K^-K^0}(s)T_{K^-K^0\to K^-K^0}(s) & g_{K^-K^0}(s)T_{K^-K^0\to\pi^-\eta^\prime}(s)\\
 g_{\pi^-\eta^\prime}(s)T_{\pi^-\eta^\prime\to\pi^-\eta}(s) & g_{\pi^-\eta^\prime}(s)T_{\pi^-\eta^\prime\to K^-K^0}(s) & 1+g_{\pi^-\eta^\prime}(s)T_{\pi^-\eta^\prime\to\pi^-\eta^\prime}(s)\\
 \end{array}
 \right) \; . \label{D matrix}
\end{equation}
Again, possible poles of eq.~(\ref{3ccmatrix}) need to be removed as explained before.

\begin{figure}[h!]
\begin{center}
\includegraphics[scale=0.75]{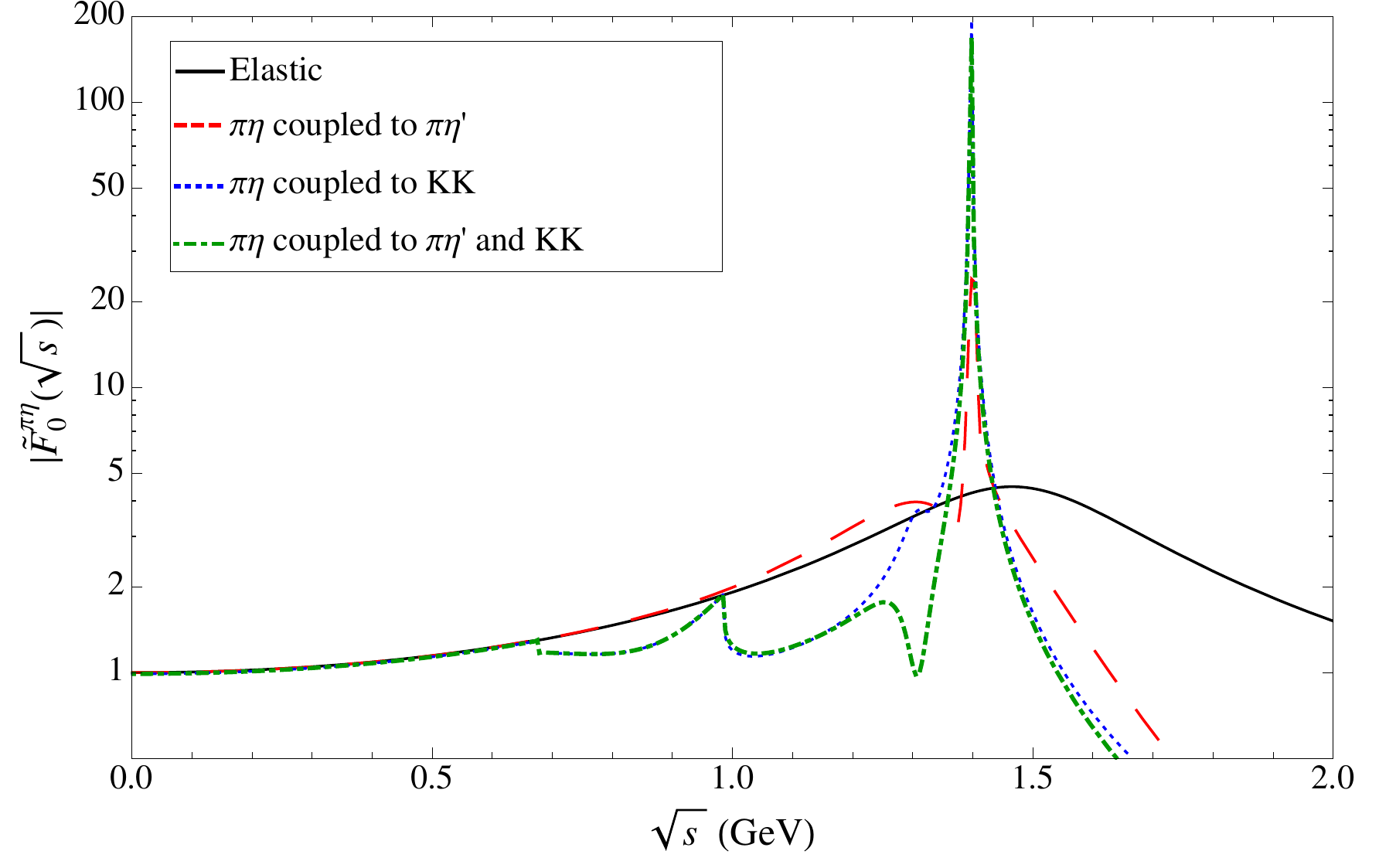}
\includegraphics[scale=0.75]{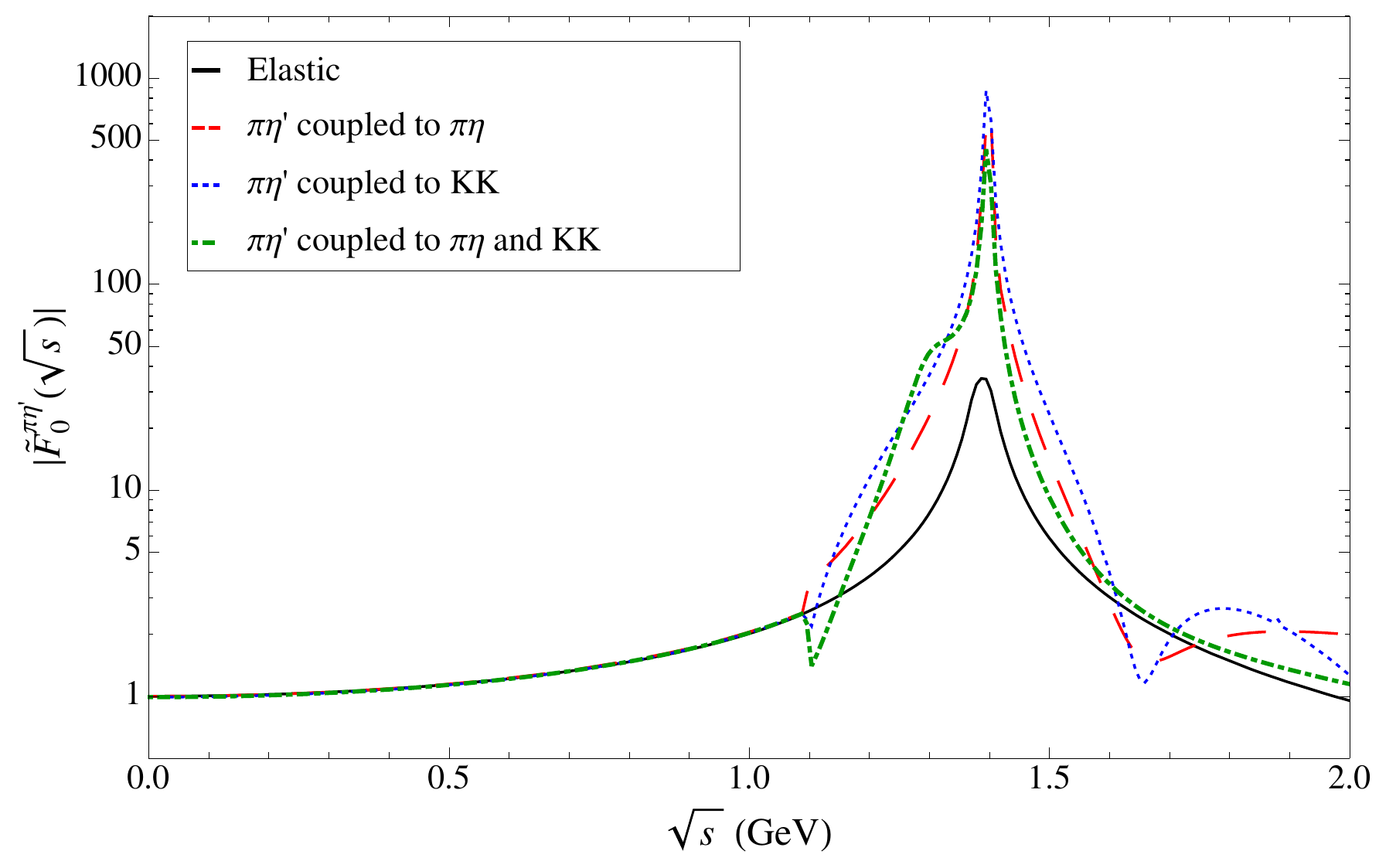}
\caption{\label{3cc} $\pi^{-}\eta$ SFF coupled to $K^{-}K^{0}$ and $\pi^{-}\eta^{\prime}$ (up) and $\pi^{-}\eta^{\prime}$ SFF coupled to $\pi^{-}\eta$ and $K^{-}K^{0}$ (down) as calculated from 
eq.~(\ref{3ccmatrix}) (green dot-dashed curve) compared to both the elastic case (black solid curve) and to the two coupled-channels cases (red dashed and blue dotted curves). All the expressions 
are normalized to unity at the origin.}
\end{center}
\end{figure}

Our results \cite{Escribano:2016ntp}, obtained by means of this unitarization procedure, are shown in figure \ref{3cc}. While for  
$\widetilde{F}_{0}^{\pi^{-}\eta}(s)$ the coupling of the $\pi^-\eta-K^-K^0$ channels gives already a good approximation to the solution of the three-coupled channels problem, this is not the case for the 
$\widetilde{F}_{0}^{\pi^{-}\eta^\prime}(s)$, where no simplified solution should be used. The complicated multi-peak-and-dip structure of these form factors is a characteristic feature of the unitarization 
of the coupled channels problem. It must be evident that a parametrization of these form factors which does not fulfill unitarity and analyticity (for instance a sum of Breit-Wigner-like functions) will 
have very little (if anything) to do with this QCD prediction.

\section{Predictions}\label{Pred}

\begin{figure}[h!]
\begin{center}
\includegraphics[scale=0.9]{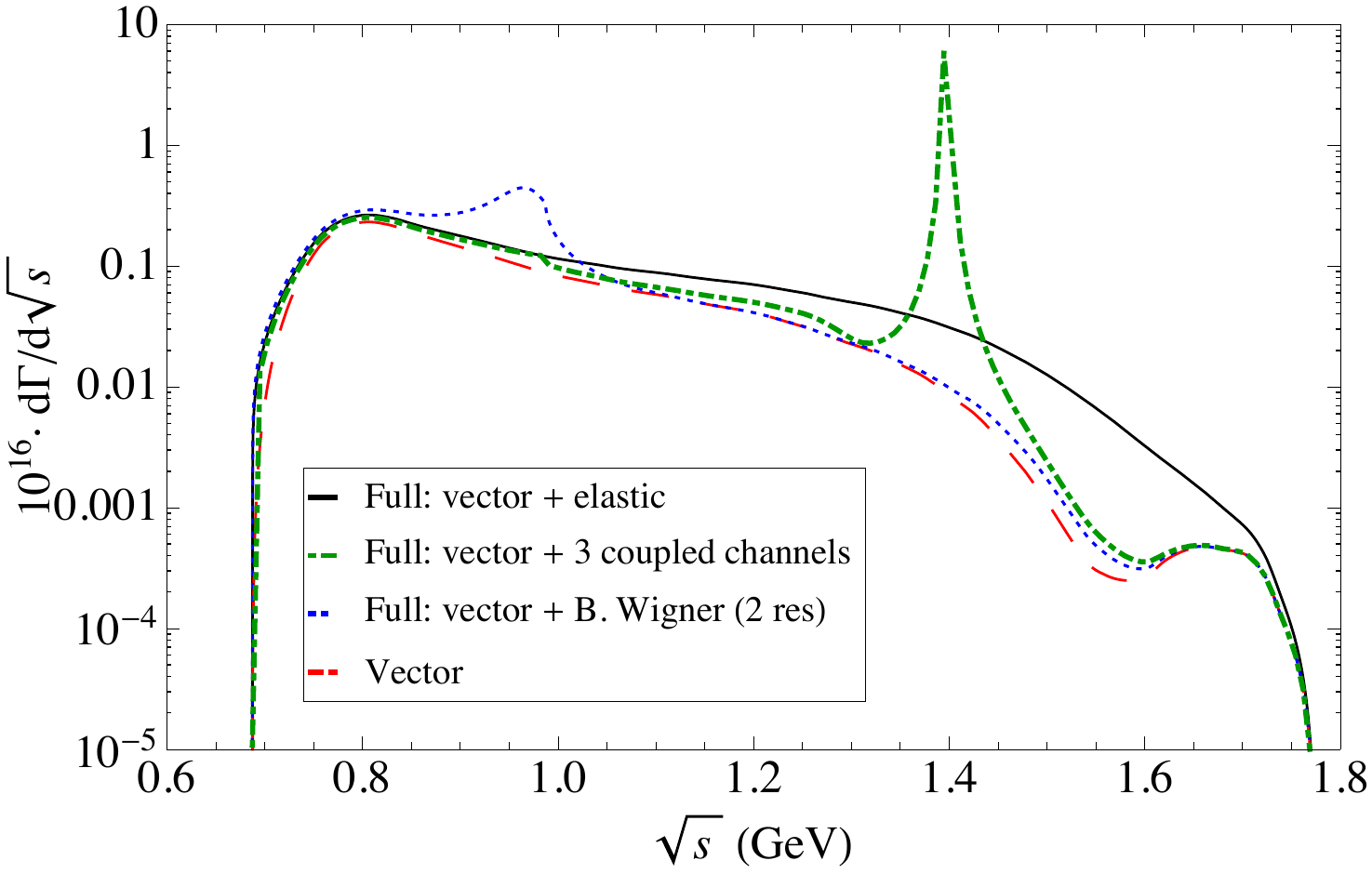}
\caption{\label{distributionpieta} Decay distribution for $\tau^{-}\to\pi^{-}\eta\nu_{\tau}$ decays. Red-dashed curve describes the contribution corresponding to the vector form factor while the other 
three curves represent the full distribution by employing the scalar form factor in its elastic version (black solid curve), in the three coupled-channels analysis (green dot-dashed curve) and using a 
Breit-Wigner formula with two resonances (blue dotted curve).}
\end{center}
\end{figure}

Our predicted spectra for the $\tau^{-}\to\pi^{-}\eta\nu_{\tau}$ decays are plotted in Fig.~\ref{distributionpieta}. As discussed previously, our reference curve is the green dot-dashed line, corresponding 
to the scalar form factor obtained solving the three-coupled channels problem. Comparison with the other curves in the figure makes clear, again, that simplified approaches should not be pursued in the 
study of these decays.

In table \ref{Tablepieta} we compare our predictions for the $\tau^{-}\to\pi^{-}\eta\nu_{\tau}$ decays to current experimental bounds and to other results in the literature. 
Earlier analyses show a spread in the vector form factor contribution that is much larger than the uncertainty resulting from its data-driven extraction that we advocate. According to 
our previous discussion it seems natural that approaches that do not require unitarity to the scalar form factor tend to underestimate its contribution to the decay width.

\begin{table*}[h!]
 \begin{center}
\begin{tabular}{|l|l|l|l|}
\hline
BR$_V\cdot10^{5}$ & BR$_S\cdot10^{5}$ & BR$\cdot10^{5}$ & Reference\cr
\hline
$0.25$& $1.60$ & $1.85$ & Tisserant, Truong \cite{Tisserant:1982fc}\cr
$0.12$ & $1.38$& $1.50$& Bram\'{o}n, Narison, Pich \cite{Bramon:1987zb}\cr
$0.15$& $1.06$ & $1.21$ & Neufeld, Rupertsberger \cite{Neufeld:1994eg}\cr
$0.36$& $1.00$ & $1.36$ & Nussinov, Soffer \cite{Nussinov:2008gx}\cr
$\left[0.2,0.6\right]$ & $\left[0.2,2.3\right]$ & $\left[0.4,2.9\right]$ & Paver, Riazuddin \cite{Paver:2010mz}\cr
$0.44$ & $0.04$ & $0.48$ & Volkov, Kostunin \cite{Volkov:2012be}\cr
$0.13$ & $0.20$& $0.33$& Descotes-Genon, Moussallam \cite{Descotes-Genon:2014tla}\cr
\hline
$0.26\pm0.02$&$1.41\pm0.09$&$1.67\pm0.09$&\textbf{Our result} \cite{Escribano:2016ntp}\cr
\hline
& & BR$\cdot10^{5}$ &Experimental collaboration\cr
\hline
&&$<14$ ($95\%$ CL)&CLEO \cite{Bartelt:1996iv}\cr
&&$<7.3$ ($90\%$ CL)&Belle \cite{Hayasaka:2009zz}\cr
&&$<9.9$ ($95\%$ CL)&BaBar \cite{delAmoSanchez:2010pc}\cr
\hline\end{tabular}
\caption{\label{BRpieta}Our branching ratio predictions for the $\tau^{-}\to\pi^-\eta\nu_{\tau}$ decays are compared to current experimental data and other authors' estimates. In the first and second 
columns the contributions from the vector and scalar form factors are displayed. The source of uncertainty arises from the errors on $\varepsilon_{\pi\eta}$ and $\varepsilon_{\pi\eta^{\prime}}$ and 
from the (uncorrelated) errors on the input values. The total branching ratio is obtained after symmetrizing and adding in quadrature all uncertainties.}
\end{center} \label{Tablepieta}
\end{table*}

\begin{figure}[h!]
\begin{center}
\includegraphics[scale=0.9]{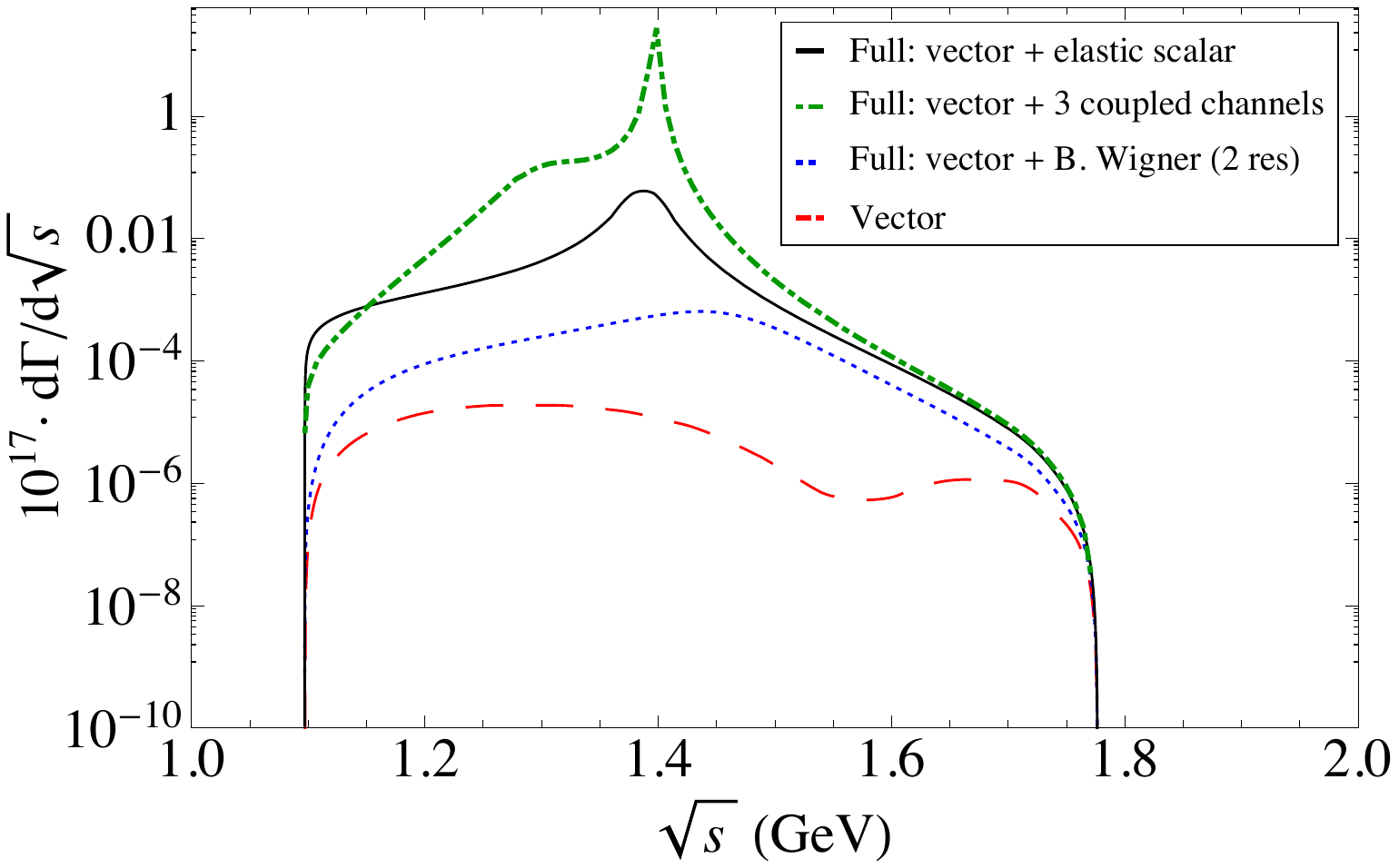}
\caption{\label{distributionpietap} Decay distribution for the $\tau^{-}\to\pi^{-}\eta^{\prime}\nu_{\tau}$ decays. Red-dashed curve describes the contribution corresponding to the vector form factor 
while the other three curves represent the full distribution by employing the scalar form factor in its elastic version (black solid curve), in the three coupled-channels analysis (green dot-dashed curve) 
and finally using the Breit-Wigner formula with two resonances (blue dotted curve).}
\end{center}
\end{figure}

Similarly, we plot our predictions for the $\tau^{-}\to\pi^{-}\eta^{\prime}\nu_{\tau}$ decays in fig.~\ref{distributionpietap}, where the reference line is the green dot-dashed curve 
corresponding to the three-coupled channels determination of the scalar form factor. It can hardly be more evident that the unitarized solution to the three-coupled channels 
problem for the relevant scalar form factor is the only meaningful description for these decays.

In table \ref{Tablepieta} we compare our predictions for the $\tau^{-}\to\pi^{-}\eta^{\prime}\nu_{\tau}$ decays to current experimental bounds and to other results in the 
literature. Again, it is obvious that simplified descriptions of the scalar form factor yield underestimated predictions for these decay rates.

\begin{table}[h!]
 \begin{center}
\begin{tabular}{|l|l|l|l|}
\hline
BR$_V$ & BR$_S$ & BR & Reference\cr
\hline
$<10^{-7}$& $\left[0.2,1.3\right]\cdot10^{-6}$ & $\left[0.2,1.4\right]\cdot10^{-6}$ & Nussinov, Soffer \cite{Nussinov:2009sn}\cr
$\left[0.14,3.4\right]\cdot10^{-8}$ & $\left[0.6,1.8\right]\cdot10^{-7}$ & $\left[0.61,2.1\right]\cdot10^{-7}$ & Paver, Riazuddin \cite{Paver:2011md}\cr
$1.11\cdot10^{-8}$&$2.63\cdot10^{-8}$&$3.74\cdot10^{-8}$ & Volkov, Kostunin \cite{Volkov:2012be}\cr
\hline
\textbf{$\left[0.3,5.7\right]\cdot10^{-10}$}&\textbf{$\left[1\cdot10^{-7},1\cdot10^{-6}\right]$}&\textbf{$\left[1\cdot10^{-7},1\cdot10^{-6}\right]$}&\textbf{Our result} \cite{Escribano:2016ntp}\cr
\hline
&& BR & Experimental collaboration\cr
\hline
&&$<4\cdot10^{-6}$ ($90\%$ CL)&BaBar \cite{Lees:2012ks}\cr
&&$<7.2\cdot10^{-6}$ ($90\%$ CL)&BaBar \cite{Aubert:2008nj}\cr
\hline
\end{tabular}
\caption{\label{pietappredictions}Our branching ratio predictions for the $\tau^{-}\to\pi^-\eta^{\prime}\nu_{\tau}$ decays are compared to current experimental bounds and other authors' estimates as it was 
done in table \ref{Tablepieta}.}
\end{center} \label{Tablepieta'}
\end{table}

In Ref.~\cite{Escribano:2016ntp} we also consider the partner $\eta^{(\prime)}\to\pi^{+}\ell^{-}\nu_{\tau}\quad (\ell=e, \mu)$ decays. Our predictions for them indicate that their discovery is out of 
reach at present and next-generation colliders. We obtain $BR(\eta\to\pi^{+}e^{-}\nu_{e}+c.c.)=0.6\cdot10^{-13}$, $BR(\eta\to\pi^{+}\mu^{-}\nu_{\mu}+c.c.)=0.4\cdot10^{-13}$ 
and $1.7\cdot10^{-17}$ for both $\eta^\prime$ decays. The first bound on any of these decays has just been set by BESIII: $\mathcal{B}(\eta\to\pi^{+}e^{-}\bar{\nu}_{e}+c.c.)<1.7\cdot10^{-4}$ and 
$\mathcal{B}(\eta^{\prime}\to\pi^{+}e^{-}\bar{\nu}_{e}+c.c.)<2.2\cdot10^{-4}$, both at the $90\%$ C.L., which are still extremely far from our predictions. The spectra of these decays 
can be found in Ref.~\cite{Escribano:2016ntp}.\\

\section{Conclusions and outlook} \label{Outlook}
According to our results, SCC should finally be discovered at Belle-II through the measurement of the $\tau^{-}\to\pi^{-}\eta\nu_{\tau}$ decays. A precision of $\sim20\%$ on both their 
measurement and the corresponding theory prediction would allow to improve the bounds on a charged Higgs \cite{Descotes-Genon:2014tla} (currently obtained from $B\to\tau\nu$). Our study aims at 
providing a QCD-based parametrization of this decay that can be useful in Belle-II searches. Upon eventual discovery, a joint analysis of meson-meson scattering amplitudes together with 
the Belle-II signal on $\tau^{-}\to\pi^{-}\eta^{(\prime)}\nu_{\tau}$ decays would yield an accurate determination of the correlations between all relevant low-energy constants and resonance 
parameters that would enable a consistency test of both sets of data in the search for non-SM contributions, i. e. \textit{genuine} SCC. In these combined fits, the usefulness of our closed-form 
solution of the multi-coupled channels problem emerges plainly.

Several things remain to be done. In BaBar and Belle-I the background from $\tau^{-}\to\pi^{-}\pi^0\eta\nu_{\tau}$ decays \cite{Inami:2008ar} limited the searches for SCC. Within a Mexican-Polish 
Collaboration \cite{Michel} we plan to include the Resonance Chiral Lagrangians prediction for these decays \cite{Dumm:2012vb} in the corresponding TAUOLA currents \cite{Shekhovtsova:2012ra} in order 
to ease the Belle-II searches. It turns out \cite{Adolfo} that some resonance contributions to the $\tau^{-}\to\pi^{-}\eta^{(\prime)}\gamma\nu_{\tau}$ decays yield sizable backgrounds to the searches 
for SCC in the corresponding non-radiative decays, and therefore need to be properly modeled in TAUOLA too, as we plan to do in the future.
\ack
I enjoyed very much collaborating with R.~Escribano and S.~Gonz\'alez-Sol\'is on the research summarized here. I thank S.~G.~S. for a careful revision of the draft of this contribution. 
Mexican funding from Conacyt and SNI (Mexico) supporting this research is acknowledged, as well as from RED-FAE to attend the XV Mexican Workshop on Particles and Fields. I congratulate 
the organizing Committee for the fruitful conference.


\begin{thebibliography}{9}
\bibitem{Michel}
  Michel Hern\'andez, a Ph. D. student co-advised by Eduard de la Cruz Burelo and Gabriel L\'opez Castro, and Jorge Mart\'inez, a postdoc of the project ``Fronteras de la Ciencia 296'', 
are involved in this project with us.
\end{thebibliography}
\end{document}